\documentclass{sig-alternate-05-2015}

\begin{document}


\setcopyright{rightsretained}
\conferenceinfo{Neu-IR'16}{SIGIR Workshop on Neural Information Retrieval, July 21, 2016, 
Pisa, Italy}

\doi{}

\isbn{}


\acmPrice{\$00.00}

%


\title{Toward Word Embedding\\ for Personalized Information Retrieval}

%
%
%
%
%

\numberofauthors{3} 
%
\author{
%
%
\alignauthor
Nawal OULD AMER \\
       \affaddr{Universit{\'e} de Grenoble }\\
       \affaddr{LIG laboratory, MRIM group Grenoble, France}\\
       \email{nawal.ould-amer@imag.fr}
\alignauthor
Philippe MULHEM \\
         \affaddr{CNRS}\\
       \affaddr{LIG laboratory, MRIM group Grenoble, France}\\
       \email{philippe.mulhem@imag.fr}
\alignauthor Mathias G{\'E}RY \\
		\affaddr{Universit{\'e} de Saint-{\'E}tienne }\\
       \affaddr{Hubert Curien Laboratory Saint-{\'E}tienne, France }\\
       \email{mathias.gery@univ-st-etienne.fr}
}

\maketitle

\begin{abstract}
This paper presents preliminary works on using \textit{Word Embedding} (\textbf{word2vec}) for query expansion in the context of Personalized Information Retrieval.
Traditionally, word embeddings are learned on a general corpus, like Wikipedia. In this work we try to personalize the word embeddings learning, by achieving the learning on the user's profile. The word embeddings are then in the same context than the user interests. Our proposal is evaluated on the CLEF Social Book Search 2016 collection. The results obtained show that some efforts should be made in the way to apply \textit{Word Embedding} in the context of Personalized Information Retrieval.
\end{abstract}

\ccsdesc[500]{Information systems~Personalization}

\printccsdesc

\keywords{Word Embedding, word2vec, Personalization, Social Book Search, Query expansion}
\section{Introduction}
\label{sec:intro}

Recent works investigate the use of \textit{Word Embedding} for enhancing IR effectiveness~\cite{AlmasriBC16,Ganguly:2015:WEB:2766462.2767780,Nalisnick:2016:IDR:2872518.2889361} or classification~\cite{kim14f}.
\textit{Word Embedding} \cite{word2vec} is the generic name of a set of NLP-related learning techniques
that seek to embed representations of words, leading to a richer representation: words are represented as vectors of more elementary components or features.
Similarly to Latent Semantic Analysis (LSA) \cite{sdx1990}, \textit{Word Embedding}
maps the words to low-dimensional (w.r.t. vocabulary size) vectors of real numbers.
For example, two vectors $\overrightarrow{t_0}$ and $\overrightarrow{t_1}$, corresponding to the words $t_0$ and $t_1$, are close in a N-dimensional space if they have similar contexts
and vice-versa, i.e. if the contexts in turn have similar words~\cite{DBLP:journals/corr/GoldbergL14}.
In this vector space of embedded words, the cosine similarity measure is classically used to identify words occurring in similar contexts.
In addition, the arithmetic operations between vectors reflect a type of semantic-composition, e.g. \emph{bass + guitar = bass guitar} \cite{dw2004}.\\

In this paper, we present an approach using \textit{Word Embedding} for Personalized Information Retrieval. The goal of our works is to provide clues about the following questions:
\begin{itemize}
\item Can \textit{Word Embedding} be used for query expansion in the context of  \textbf{social collection} ?
\item Can \textit{Word Embedding} be used to \textbf{personalize} query expansion?
\end{itemize}

The first question is motivated by our participation in the CLEF Social Book Search task in 2016 (similar to the Social Book Search task in 2015~\cite{Koolen2015}). Our concern was related to the fact that the topics provided contain non-topical terms that may impact the usage of \textit{Word Embedding}.\\

The second question tackles more specifically the usage of \textit{Word Embedding} when the learning is based on the user's profile in order to select personalized words for query expansion. The idea is to select words that occur in the same context as the terms of the query. We compare then \textit{Word Embedding} learned on the whole collection of Social Book Search, called the \textit{Non Personalized Query Expansion}, \textit{versus} \textit{Word Embedding} learned on the user's profiles, called the \textit{Personalized Query Expansion}.\\

The paper is organized as follows: Section~\ref{sec:proposal} presents the proposed approach.
The experiments on the official CLEF Social Book Search collection are presented and discussed in Section~\ref{sec:expes}, and the results are commented in Section~\ref{sec:results}. We conclude this work in Section~\ref{sec:conclusion}.

\section{Personalized Query Expansion}
\label{sec:proposal}
\subsection{User Modeling}
\label{sec:usermodeling}
In the context of Social Book Search, each user is represented by his \textit{catalog} (i.e. set of books) and other information such as tags and the ratings that he assigns to the books~\cite{Koolen2015}. All these information describe the interests of the user.\\

We represent a user $u$ as one document $d_{u}$ which is the concatenation of all the documents present in his catalog. The profile of $u$, noted $p_u$, is then represented by the set of words in $d_u$:
\begin{equation}
p_u = \{w_1, w_2, w_3, ..., w_n\}
\end{equation}

 \subsection{Term Filtering}
 \label{Term Filtering}
As stated before, \textit{Word Embedding} can potentially be helpful in the selection of terms related to the query. Usually, embedded terms are used for query expansion, as extensions of one of the query term.
Despite the effectiveness of \textit{Word Embedding} to select embedded terms, it could happen that their use for query expansion decreases the effectiveness of the system.\\

In fact, if an extended term is a noisy term (because of its ambiguity for instance, or because it is not a topical term of the query), then the set of its resulting word embeddings will increase the non-topical noise in the extended query. For example, in the queries (taken from the topics of Social Book Search 2016) \textit{``Help! I Need more books"}, \textit{``New releases from authors that literally make you swoon..."} or \textit{``Favorite Christmas Books to read to young children"}, the majority of theses words are not useful for expansion, like \textit{``new, good, recommend, make, you, etc."}. Therefore, these words need to be filtered out of the queries before the expansion process.
We chose to remove all the adjectives words from the queries. To do that, we use an English stop-adjective list in addition to the standard English stop-list.\\

Then, from a user query $q = \{ t_1,t_2, ... , t_m \}$, we note $q_f = \{ t_1,t_2, ... , t_o \}$ the filtered query.

\subsection{Word Embedding Selection}
\label{sec:wordembeddingselection}
Once we have a filtered query $q_f$ as described above (cf. subsection \ref{Term Filtering}), we select the top-k word embeddings to be used as extensions. This selection is achieved in three steps for each term $t$ of the filtered query $q_f$:
\begin{itemize}
\item[\textbf{i)}] Building a set of word  embeddings for $t$ using the cosine similarity between $t$ and all words in the training corpus; 
\item[\textbf{ii)}]Filtering out from the set of word  embeddings of \textbf{i)} the terms that have the same stem than $t$ using English Porter Stemmer, in order to avoid overemphasizing the variations of $t$; 
\item[\textbf{iii)}] Selecting the top-k word embeddings of $t$ from the filtered set of word embeddings of \textbf{ii)}. 
\end{itemize}

Then, the output of the selection of word embeddings is: 
\begin{equation}
\label{matrice_wordembedding}
WordEmbedding_{w2v}(q_f) = {\left\lbrace
\begin{array}{l}
em\_t_{11}, em\_t_{12}, ..., em\_t_{1k},  \\
em\_t_{21}, em\_t_{22}, ..., em\_t_{2k}, \\
... \\
em\_t_{o1}, em\_t_{m2}, ..., em\_t_{ok}  
\end{array}
\right\rbrace
}
\end{equation}

Where $WordEmbedding_{w2v}(q_f)$ denotes the function that returns a set of word embedding for a given filtered query $q_f$, and $em\_t_{ij}$ denotes the $j^{th}$ element of the top-k word embeddings of $t
_i$.

\subsection{Ranking Model}
\label{sec:resutranking}
The final expanded query $q_{new}$ is the union of the original user query $q$ and the word embeddings set as follow:
\begin{equation}
q_{new} = q \cup WordEmbedding_{w2v}(q_f)
\end{equation}

The score for each document $d$ according to the expanded query $q_{new}$ and the user $u$ is computed according to a classical Language Model with Dirichlet smoothing.

\section{EXPERIMENTS}
\label{sec:expes}
\subsection{Dataset}
Our experiments are conducted on Social Book Search dataset~\cite{Koolen2015}.
\begin{itemize}
\item Documents: The documents collection consists of 2.8 millions of books descriptions with meta-data from Amazon and LibraryThing. Each document is represented by book-title, author, publisher, publication year, library classification codes and user-generated content in the form of user ratings and reviews.\footnote{http://social-book-search.humanities.uva.nl/\#/suggestion}
\item Users: The collection provides profiles of 120 users. Each user is described by his catalog (i.e. a set of books), tags, and rating.
\item Queries: The collection contains 120 user queries. Due to the nature of the queries, we chose 28 queries for our experiments: these queries have topical content (i.e. a classical IR system has then chances to get relevant results) and the profile of the query issuer is not empty.

\end{itemize}

\subsection{Learning of Word Embedding\label{sec:wordembedding}}
Here, we describe the process of learning word vectors (see section~\ref{sec:intro}). We use two training sets. 
\begin{itemize}
\item The first one is called \textit{\textbf{The Non Personalized Corpus}}: the train is built on the whole Social Book Search Corpus. 
\item The second one is called \textit{\textbf{The Personalized Corpus}}: we build a personalized train corpus for each user. 
\end{itemize}
The training process is the same for the two corpora:
\begin{enumerate}
\item  \textit{\textbf{Non Personalized Corpus train process:}}\\
We train \textbf{word2vec}~\cite{word2vec} on the Social Book Search corpus. \textbf{word2vec} represents each word $w$ of the training set as a vector of features, where this vector is supposed to capture the contexts in which $w$ appears. Therefore, we chose the Social Book Search corpus to be the training set, as the training set is expected to be consistent with the data on which we want to test.
To construct the training set from the corpus, we simply concatenate the content of all the documents, without any particular pre-processing such as stemming or stop-words removal, except some text normalization and cleaning operations such as lower-case normalization, removing HTML tags, etc. The concatenated document is the input training set of \textbf{word2vec}. The size of the training set is $\sim12.5 GB$; it contains $2.3$ billions words, and the vocabulary size is about $600K$.\\

The training parameters of \textbf{word2vec} are set as follows:
\begin{itemize}
	\item continuous bag of words model instead of the skip-gram (\textbf{word2vec} options: cbow=1);
    \item the output vectors size or the number of features of resulting word-vectors is set to 500 (\textbf{word2vec} options: size=500);
    \item the width of the word-context window is set to 8 (\textbf{word2vec} options: window=8);
    \item the number of negative samples is set to 25 (\textbf{word2vec} options: negative=25).
\end{itemize}

\item \textit{\textbf{Personalized Corpus train process:}}\\
For this corpus, each user is described as a document $d_u$ (containing the documents belonging to his catalog, cf. Section \ref{sec:usermodeling}) and we train \textbf{word2vec} on each user document $d_u$ using the same process than the \textit{Non Personalized Corpus} train process.
\end{enumerate}
\subsection{Parameters and Tested Configurations}
The ranking model is achieved using the Language Model with Dirichlet smoothing. All documents are retrieved using Terrier search engine \cite{ounis06terrier-osir} with $\mu =50$.\\

We compare the following variations: 
\begin{enumerate}
\item \textbf{Query filtering:} the query is filtered or not, by removing the stop-word and the adjectives as stated in section \ref{Term Filtering};
\item \textbf{Query expansion:} the query is expanded or not using the \textit{Word Embedding} function;
\item \textbf{Personalization:} the query is personalized (using the \textit{Personalized Corpus}) or not (using the \textit{Non Personalized Corpus}).
\end{enumerate}

The tested configurations are presented in Table~\ref{tab:configurations}.

\begin{table}[htb]  
\label{tab:configurations}
\centering
\begin{tabular}{|c||c|c|} \hline
Configuration & Query filtering & Expansion\\ 
\hline
\hline
Conf$_1$ (baseline) & Original ($q$) & -\\ \hline
Conf$_2$ & Filtered ($q_f$) & - \\ \hline
Conf$_3$ & Filtered ($q_f$) & Non Personalized \\ \hline
Conf$_4$ & Filtered ($q_f$) & Personalized \\ \hline
Conf$_5$ & Original ($q$) & Non Personalized \\ \hline
Conf$_6$ & Original ($q$) & Personalized \\ \hline
\end{tabular}
\caption{Tested configurations}
\end{table}

\section{RESULTS}
\label{sec:results}
In this section, we present and comment the results of the above configurations. All of these configurations lead to quite low MAP values, but this is consistent with the official CLEF Social Book Search results.

\subsection{Query Filtering}

Table~\ref{filteredornot} reports the Mean Average Precision (MAP), Mean Reciprocal Rank (MRR), and precision at 10 documents (P@10) evaluation measures obtained with the configuration Conf$_1$ (without query filtering and without query expansion) and the configuration Conf$_2$ (with query filtering and without query expansion). As we can see, filtering the query terms with the stop-adjective list improves the MAP values, but surprisingly nor the MRR neither the P@10. The choice of filtering the adjectives may be not the most effective way to deal with noisy terms.

\begin{table}[htb]  
\label{filteredornot}
\centering
\begin{tabular}{|c||c|c|c|} \hline
Configuration &MAP & MRR & P@10\\ 
\hline
\hline
Conf$_1$ ($q$, No QE) & 0.0266& 0.1478 & 0.0464\\  \hline
Conf$_2$ ($q_f$, No QE) & 0.0309 & 0.1436 & 0.0393\\
\hline\end{tabular}
\caption{With vs without query filtering}
\end{table} 

\subsection{Personalized Query Expansion (with filtering)}

\begin{figure}[htb]   
        \includegraphics[width=8.5cm]{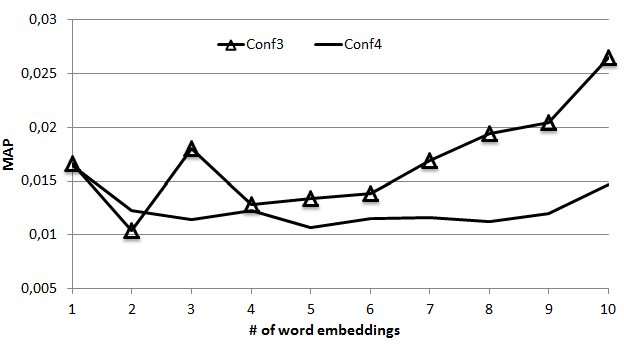}
    \caption{Non Personalized vs Personalized Query Expansion (with filtering)}
    \label{Personalized Query Expansion (with filtering)}
\end{figure}

Figure~\ref{Personalized Query Expansion (with filtering)} reports the MAP effectiveness over the number of word embeddings. As we can see, in most of the cases the non personalized approach outperforms the personalized approach. The personalized approach shows a better result only when the word embeddings are limited to the top-2 terms. We also remark that both the two approaches underperfom the filtered non expanded approach Conf$_2$, which shows that inadequate words are added. For $Conf_3$, we see that adding more that 8 terms lead to better results, which is not really the case to the personalized configuration $Conf_4$ where the MAP evolution is quite flat. In this case, no added term seems to play a positive role.

\subsection{Personalized Query Expansion (without filtering)}

\begin{figure}[htb]   
        \includegraphics[width=8.5cm]{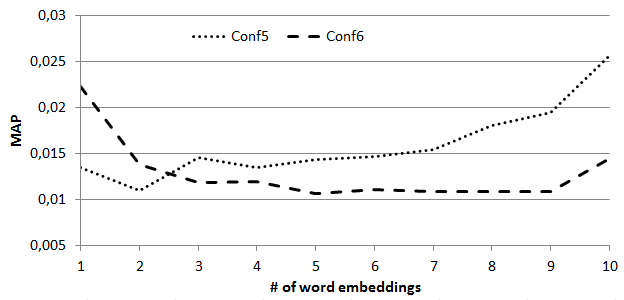}
    \caption{Personalized Query Expansion (without filtering)}
    \label{Expended Query Not Personalized vs Expended Query Personalized}
\end{figure}

The figure~\ref{Expended Query Not Personalized vs Expended Query Personalized} reports the MAP value evolution over the number of word embeddings for the configurations Conf$_5$ (non personalized query expansion, without filtering) and the Conf$_6$ (personalized query expansion, without filtering). As we can see, in most of the cases, the non personalized approach outperforms the personalized one.
However, these two curves in figure~\ref{Expended Query Not Personalized vs Expended Query Personalized} behave quite differently: the personalized approach is better only when adding the top-1 and top-2 words, leading to consider that the first two personalized terms are interesting, whereas the non personalized expansion behaves better and better as the number of terms increases, leading to consider that ``good" terms (according to the user) appear in the non personalized case.

\subsection{Discussion}
\begin{figure}[htb]   
        \includegraphics[width=8.5cm]{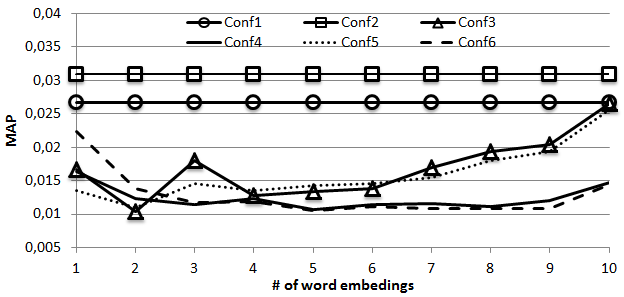}
    \caption{All Configurations together}
    \label{all}
\end{figure}

The figure \ref{all} reports the MAP value over the number of word embeddings for the six configurations. As we can see, the results of the configurations: 1) Conf$_3$ and Conf$_5$, and 2) Conf$_4$ and Conf$_6$ are quite similar. \\

We observe that the configurations with query expansion (Conf$_1$ and Conf$_2$) have still the best results, and outperform all the configurations with query expansion (i.e.  Conf$_3$, Conf$_4$, Conf$_5$ and Conf$_6$).\\

As presented above, in most of the cases, the expanded approach fails to improve results.
We may explain these results by the quality of the corpus. In fact, the documents collection describes the reviews of users for books. Therefore, it's still difficult to extract the context of terms and select the similar terms in the similar context.\\

The personalized \textit{Word Embedding} fails to improve the results comparing to any configurations. We first can explain the results by the same quality problem than for the documents collection. In fact, the user is represented by the documents that appear in his catalog. So, these documents present the reviews of the user about the books. Secondly, the learning of \textbf{word2vec} is effective if a large amount of data is available. However, the users' profiles correspond to short documents. Therefore, the amount of learning data may not reach the limit under which no convergence is possible for \textbf{word2vec}.

\section{Conclusion}
\label{sec:conclusion}
The focus of this paper was to study the integration of word embeddings for the Social Book Suggestion task of CLEF 2016, according to non-personalized and to personalized query expansion.\\

We found that the nature of the queries poses a great challenge to an effective use of \textit{Word Embedding} in this context.
Future works may help to better understand this behavior.\\

The second point is related to the fact that the personalization using word embeddings did not lead to good results. The reasons could be multiple: the quality of the description of the user's profiles, the lack of data that we get to describe the profile which does not allow word embeddings to be used effectively. Here also, future works may help to find solutions to these problems.
Especially, is it possible to counteract this limitation by adding other inputs (user neighbors for instance)? This question is largely open for now.

\bibliographystyle{abbrv}
\bibliography{sigproc}  
\end{document}